\documentclass{article}
\usepackage[a4paper, margin=1in]{geometry}
\usepackage{cite}
\usepackage{lettrine}
\usepackage[pdftex]{graphicx}
\usepackage{mdwmath}
\usepackage{mdwtab}
\usepackage{amssymb}
\usepackage[cmex10]{amsmath}
\usepackage{eqparbox}
\usepackage{array}

\newcommand\norm[1]{\lVert#1\rVert}

\usepackage{algorithmic}
\usepackage{varwidth}
\usepackage{multirow} 
\usepackage{tabularx}
\usepackage{array}
\usepackage{caption}
\usepackage{subcaption}
\usepackage{wrapfig}
\usepackage{lscape}
\usepackage{rotating}
\usepackage{epstopdf}

\usepackage{varwidth}
\usepackage{tikz}
\usepackage{amssymb}
\usepackage{amsmath}
\usetikzlibrary{shapes,arrows,fit,calc}
\pgfdeclarelayer{back}
\pgfsetlayers{back,main}
\usepackage{xurl}
\usepackage[mode=buildnew]{standalone}
\usepackage{pdfpages}
\usepackage{authblk}
\providecommand{\keywords}[1]
{
  \small	
  \textbf{\textit{Keywords---}} #1
}

\title{A Quantum Fuzzy-based Approach for Real-Time Detection of Solar Coronal Holes}
\author[1]{Sanmoy Bandyopadhyay}
\author[2]{Suman Kundu}
\affil[1,2]{ Department of Computer Science and Engineering, Indian Institute of Technology, Jodhpur, Karwar, 342030 India}
\affil[1]{\textit {sanmoy1985@rediffmail.com}}
\affil[2]{\textit {suman@iitj.ac.in}}
\date{}

\begin{document}

\maketitle

\begin{abstract}
The detection and analysis of the solar coronal holes (CHs) is an important field of study in the domain of solar physics. Mainly, it is required for the proper prediction of the geomagnetic storms which directly or indirectly affect various space and ground-based systems. For the detection of CHs till date, the solar scientist depends on manual hand-drawn approaches. However, with the advancement of image processing technologies, some automated image segmentation methods have been used for the detection of CHs. In-spite of this, fast and accurate detection of CHs are till a major issues. Here in this work, a novel quantum computing-based fast fuzzy c-mean technique has been developed for fast detection of the CHs region. The task has been carried out in two stages, in first stage the solar image has been segmented using a quantum computing based fast fuzzy c-mean (QCFFCM) and in the later stage the CHs has been extracted out from the segmented image based on image morphological operation. In the work, quantum computing has been used to optimize the cost function of the fast fuzzy c-mean (FFCM) algorithm, where quantum approximate optimization algorithm (QAOA) has been used to optimize the quadratic part of the cost function. The proposed method has been tested for 193 \AA{} SDO/AIA full-disk solar image datasets and has been compared with the existing techniques. The outcome shows the comparable performance of the proposed method with the existing one within a very lesser time.  
\end{abstract}


\section{Introduction}\label{Section-1}
\lettrine{T}{he} detection of solar coronal holes (CHs) is an important problem in space weather analysis. CHs are the visible dark region of extreme-ultra violet (EUV) and X-ray images of solar corona. CHs are one of the major source of high-speed solar wind streams (HSSS) which results in severe geomagnetic storms. Geomagnetic storms have an acute impact on ground infrastructures such as communication devices, and the electrical power supply at the polar region as-well-as space-based infrastructure like low earth-orbiting, communication and navigation satellites, .

As a consequence, there rise the demand for proper analysis of the solar CHs for prediction of the geomagnetic storm and its related phenomenon. These CHs regions on the Sun surface can be visualized as a dark region only in space based extreme-ultra violet (EUV) and X-ray images; that cannot be visible by the ground based observatories as these wavelengths are blocked by the Earth atmosphere. The regions are open unipolar magnetic field on the solar surface, having lower temperature compared to surrounding region of the solar corona and appear dark due to the low density of plasma in contrast to the surrounding corona region. 
The study of CHs has been expedite with the launch of space-based solar observatories such as SOlar and Heliospheric Observatory (SOHO), SOLar Orbiter (SOLO), Solar Dynamics Observatory (SDO), Solar Terrestrial Relations Observatory (STEREO). With the introduction of these solar observatories, there has been a huge increase in the volume, velocity, variety, value and veracity of the solar data that has induced the epoch of Big Data in the domain of solar studies. These data will increase more in number in coming future with the introduction of Aditya-L1, India's first solar observational satellite. The solar images captured using these observatories are characterized by multi-scale, multi-wavelength and multi-source. 

In the midst of these solar observatories, SDO launched in 2010, provides detailed solar view that was never possible before. It provides ultra-high-definition images of the Sun in 13 various wavelengths by means of two on-board image capturing devices, namely; Atmospheric Imaging Assembly (AIA) and Helioseismic and Magnetic Imager (HMI) \cite{2012SoPh..275....3P}. Each of the wavelengths takes after a particular locale of the heliosphere, which covers the parcels from the Sun's surface to the upper comes of the solar corona \footnote{\url{https://www.nasa.gov/mission_pages/sdo/spacecraft/index.html}}. SDO is one of the biggest and wealthiest Sun's image information storehouses, which capture approximately 70000 Sun's images per day \cite{Kucuk2017}. The solar image seizing rate also varies for these observatories. SDO seizes 1 image each second. SDO contains images of 10 diverse wavelengths, measured in angstroms \AA, seized with its AIA device module at an interim of each ten to twelve seconds \cite{Kucuk2017}. AIA by and large take images in multiple ultraviolet (UV) and EUV pass bands \cite{Lemen2012}. SDO's AIA instrument (SDO/AIA) has image resolution of resolution of $ 4096\times4096 $, which is twice the image resolution of STEREO (i.e., $ 2048\times 2048 $) and 4 times larger imaging resolution than SOHO (i.e., $ 1024\times 1024 $). 

To cope up these big incoming solar data and to perform real-time analysis of the solar CHs, there rise the necessity for the development of proper and fast image segmentation algorithm. With the development of image processing techniques some of the image segmentation methodologies such as; region-growing based strategy \cite{Caplan_2016}, convolution neural network (CNN) \cite{illarionov2018segmentation, refId1, 8855114}, intensity thresholding \cite{Krista2009}, spatial possibilistic clustering algorithm (SPoCA) \cite{verbeeck2014spoca}, Chan-Vese active contour model (C-VACM) \cite{902291, Boucheron2016}, parameterized online region-based active contour method (POR-ACM) \cite{Bandyopadhyay2020, 9113512}, fast fuzzy c-mean followed by static contour model (FFCM-SCM) ) \cite{Bandyopadhyay2020, 10.1007/978-981-15-8366-7_5}, fuzzy energy-based dual contour model (FEDCM) \cite{BANDYOPADHYAY20202435} have been introduced. But none of the technique is capable to produce CHs segmentation results within real-time. 

Thus to overcome the mentioned drawbacks, in this paper, we proposed a quantum computing based fast fuzzy c-mean (QCFFCM) methodology, where quantum optimization is used to optimize the energy function of the fast fuzzy c-mean (FFCM) model to segment out the solar image. We used quantum inspired alternating direction method of multipliers (ADMM) for the energy function minimization purpose. Later, area based image morphological operation has been used for extracting out the CHs region from the segmented image. The proposed algorithm is seen to work faster even for simulated quantum environment as compared to the traditional methods of CHs detection. The main contribution of the work can be summarized as
\begin{itemize}
    \item A novel technique of Quantum-FFCM for CHs detection. ADMM is used for quantum optimization with the FFCM algorithm. To our best knowledge this is the first approach to bring quantum algorithmic technique for solar CHs detection problem.
    \item Demonstrating the applicability of the technique on large number of solar CHs images, specifically 365 images from the year 2017, with quantitative analysis.
\end{itemize}
 
The paper is organized as follows: Section \ref{Section-3} highlight the details about the proposed methodology. Experiments and results are reported in Section \ref{Section-4}. Finally, Section \ref{Section-5} provides the conclusions with discussions.

\section{Related Work}\label{Section-2}
National Oceanic and Atmospheric Administration's Space Weather Prediction Center (NOAA/SWPC) provides hand-drawn (HD) synoptic maps indicating broad outline of the solar CHs regions for each day\footnote{\url{https://www.swpc.noaa.gov/products/solar-synoptic-map}}. Harvey and Recely \cite{Harvey2002} had generated daily computer-assisted HD CHs images using He I 10830 \AA{} spectroheliograms (from the Kitt Peak Vacuum Telescope (KPVT)) and photospheric magnetograms. With the development of first automated CHs detection technique utilizing morphological image analysis, thresholding and smoothing by Henney and Harvey \cite{2005ASPC..346..261H}, few more image segmentation methodologies have been introduced for CHs detection. Segmentation technique such as intensity thresholding method had been applied for the CHs detection task in solar image \cite{Krista2009}. In spite of success in detection of CHs region in solar image, the method fail to attend the accuracy level of the segmentation results due to incorrect boundary selection \cite{Krista2009}. Region growing and histogram based segmentation technique had been applied for solar image feature extraction and afterward combination of both region based and edge based techniques had been introduced for extraction of RoI from the corresponding images \cite{Scholl2009}. The algorithm requires visually varient features for CHs detection more accurately. Caplan \emph{et al.} \cite{Caplan_2016} had too introduced region-growing based strategy for CHs extraction. Accuracy of the output generated using this method relies on proper selection of the seed points in images. Watershed segmentation method had been used by Ciecholewski \cite{CIECHOLEWSKI2015203} for the detection of CHs in solar image. Although the method succeeded in segmenting out the CHs from the solar image, however the method often results in over-segment the image. Modified SPoCA suite for detection of active region (AR), coronal hole (CH) and quiet Sun (QS) regions in solar image had been implemented by Verbeeck \emph{et al.} \cite{verbeeck2014spoca}. Due to the utilization of three separate fuzzy clustering algorithms, the method is having numerical complexity. Illarionov and Tlatov \cite{illarionov2018segmentation} had implemented CNN based method for solar CHs detection. Though the technique is capable of detecting CHs, however, it requires a large database of solar images to train and validate the network. Thus, the time requirement for generation of first result is quite high for a single image. To reduce the time complexity the original size of the solar image had been down sampled to $ 256 \times 256 $ resolution. C-VACM \cite{902291} had been used by Boucheron \emph{et al.} \cite{Boucheron2016} for the extraction of CHs from the solar image. The contour initialization is a major issue in this type of segmentation model. Moreover during the process of CHs extraction the technique also extract out the solar filament channels. To overcome the presence of these region additional HMI magnetogram data has been used.   

Alongside, these segmentation methods alone are not sufficient to carry forward the task of solar CHs detection. The techniques require additional preprocessing algorithm to remove the solar limbs before segmenting the solar image \cite{Bandyopadhyay2020,BANDYOPADHYAY20202435}. For this purpose the algorithms need additional information from the metadata of the solar image, specially the information about solar disk radius length and co-ordinate points of solar disk center for elimination of solar limbs. As such, most of these stated techniques are supervised one.  

Bandyopadhyay \emph{et al.} \cite{Bandyopadhyay2020,BANDYOPADHYAY20202435,9113512} had introduced fully automated contour-based CHs segmentation technique, where no metadata had been used to eliminate the limbs regions from the solar image. Basically, in the introduced techniques CHT \cite{Duda:1972:UHT:361237.361242} has been used to initialize contour for the contour-based technique which eliminate the solar limbs during the segmentation process. In \cite{Bandyopadhyay2020} and \cite{9113512}, POR-ACM had been used for solar CHs segmentation. Although the method does not require extra information from the metadata of solar image for CHs segmentation, but, the execution time is little bit higher than the other existing technique. To reduce the execution time for detection of CHs a FFCM-SCM had been introduced by Bandyopadhyay \emph{et al.} \cite{Bandyopadhyay2020, 10.1007/978-981-15-8366-7_5}. In the work, first fast fuzzy c-mean algorithm \cite{Anton2019} had been used for solar image segmentation, then CHT inspired static contour model had been used to extract out the CHs regions from the segmented image \cite{Bandyopadhyay2020}. However, both POR-ACM and FFCM-SCM have the limitation that they detect the redundant filament regions during CHs segmentation. To eliminate these filament regions image morphological opening and closing operation had been used, which bypass the use of HMI solar data to eliminate the redundant filament region as used in \cite{Boucheron2016}. However, the methods unable to eliminate all the filament regions from the image. Alongside, it eliminate the filament regions at the cost of total CHs area. To overcome this drawback and to combine the advantage of two methods namely; POR-ACM and FFCM-SCM, in \cite{BANDYOPADHYAY20202435}, FEDCM had been developed, where fuzzy energy-based active contour model (FEACM) had been used for the segmentation of solar disk image later fuzzy energy-based static contour model had been implemented to extract out CHs regions from the segmented image. In the work morphological area operation had been used to eliminate the redundant filament region \cite{BANDYOPADHYAY20202435}.

\section{Proposed Work}\label{Section-3}
In order to speed up the detection of the CHs regions in a solar image in this work a quantum computing inspired FFCM has been proposed. The proposed model mainly composed of two algorithms these are: FFCM, that has been used to fasten the clustering of the solar image, while the quantum computing technique has been used to find the optimal clusters center. FFCM introduced by Semechko \cite{Anton2019} uses the histogram of the image intensities instead of using individual pixel's intensities of the raw image data during the clustering process as done in classical fuzzy c-means clustering (FCM). Thus, this in turn help in achieving the computational efficiency. The overall energy function for the classical FCM algorithm is given by equation \eqref{1} \cite{BEZDEK1984191};
\begin{equation}\label{1}
{\scriptsize
J_{E}=\sum_{j=1}^{N}\sum_{i=1}^{c}{\mu_{ij}}^{m}\norm{x_{j}-\upsilon_{i}}^{2}.}
\end{equation}         
Where, $ \mu_{ij} $ denotes the membership function of the $ j^{th} $ pixel in $ i^{th} $ cluster with intensity level of $ x_{j} $. The total number of pixels in the image and total cluster number are represented by $ N $ and $ c $ respectively. The term $ m $ in equation \eqref{1} represent the fuzziness constant and the term $ \upsilon_{i} $ indicate the $ i^{th} $ cluster center, which is represented by the equation \eqref{2} as follow:  
\begin{equation}\label{2}
{\scriptsize
\upsilon_{i}=\frac{\sum_{j=1}^{N}{\mu_{ij}}^{m} x_{j}}{\sum_{j=1}^{N}{\mu_{ij}}^{m}};}
\end{equation}
where the membership function $ \mu_{ij} $ is recalculated using,
\begin{equation}\label{3}
{\scriptsize
\mu_{ij}=\left[ \sum_{k=1}^{c}\left( \dfrac{\norm{x_{j}-\upsilon_{i}}}{\norm{  x_{j}-\upsilon_{k}}}\right) \right]^{-\frac{1}{m-1}},}
\end{equation}
which is initially randomly assigned.

Now in this case the number of image pixels, $ N $ will be quite large when dealing with solar disk image of size $ 4096 \times 4096 $. As a consequence, the convergent rate of the calculation will be quite slow \cite{4637446,li2007modified}. Thus to overcome the drawback of the classical FCM, in FFCM, the raw data term $ x_{j} $ has been replaced by the histogram space $ n_{L}L $. Here the term $ n_{L} $ denotes the frequency of the occurrence of pixels with intensity value $ L $ represented by;
\begin{equation}\label{4}
\begin{gathered}
    {\scriptsize
    n_{L}=\sum_{j=0}^{N}\delta[x_{j}-L],}\\
    {\scriptsize
    L\in \{0,1,. . .,255 \},}\\
    {\scriptsize
	\delta[x_{j}-L] =}
	\begin{cases}
	{\scriptsize
	1 \;\;\text{$ x_{j}=L $}}\\
	{\scriptsize
	0 \;\;\text{$ x_{j}\neq L $}.}
	\end{cases} 
\end{gathered}  
\end{equation}
Thus, the overall energy function for FFCM is given by;
\begin{equation}\label{5}
{\scriptsize
J_{E}=\sum_{L=0}^{255}\sum_{i=1}^{c}{\mu_{iL}}^{m}L\norm{n_{L}-\upsilon_{i}}^{2}.}
\end{equation}
As such, equations \eqref{2} and \eqref{3} can be rewritten as equations \eqref{6} and \eqref{7};
\begin{equation}\label{6}
{\scriptsize
\upsilon_{i}=\frac{\sum_{L=0}^{255}{\mu_{iL}}^{m} n_{l}L}{\sum_{L=0}^{255}{\mu_{iL}}^{m}};}
\end{equation}
where, $ \mu_{iL} $ represents the modified membership value for $ L^{th} $ intensity pixel in $ i^{th} $ which is denoted by,
\begin{equation}\label{7}
{\scriptsize
\mu_{iL}=\left[ \sum_{k=1}^{c}\left( \dfrac{\norm{L-\upsilon_{i}}}{\norm{  L-\upsilon_{k}}}\right) \right]^{-\frac{1}{m-1}}.}
\end{equation}
So, the searching space reduced to 255 instead of $ N $, which in turn will fasten the clustering process. However, in both the cases FCM converges when;
\begin{equation}\label{8}
{\scriptsize
\norm{\hat{\mu}^{k+1}-\hat{\mu}^{k}}<\epsilon,}
\end{equation}
where, $ k $ denotes the step of iteration and $ \epsilon $ represents the threshold value which generally lies between 0 and 1. As such, the accuracy of the clustering depends on specification of the $ \epsilon $ value. Moreover, lower the $ \epsilon $ value the better clustering result can be achieved but at the cost of more numbers of iteration.

So to surpass this issue and to find the optimal clusters center in this work the energy function given in equation \eqref{5} have been optimized using quantum driven multiblock ADMM heuristic (M-ADMM-H) algorithm. M-ADMM-H has the ability to solve “quadratic plus
convex” mixed binary optimization (MBO) problems on classical computers as-well-as on current noisy quantum devices whenever a quadratic unconstrained binary optimization (QUBO) solver is available \cite{2020}. Here in the work a three-block implementation of ADMM heuristic (3-ADMM-H) has been used for the optimization purpose as it is capable to finds solutions of higher quality in case the continuous decision variables as-well-as guaranteed to converges to feasible and optimal solutions \cite{2020}. The model optimize equation \eqref{1} in the form,
\begin{subequations}\label{9}
\begin{equation}\label{9a}
{\scriptsize
\operatorname*{minimize~~}_\mu \sum_{j=1}^{N}\sum_{i=1}^{c}{\mu_{ij}}^{m}\norm{x_{j}-\upsilon_{i}}^{2}}
\end{equation}
\begin{equation}\label{9b}
\begin{gathered}
{\scriptsize
    \operatorname*{subject~to:~~} \sum_{j=1}^{N}\sum_{i=1}^{c}{\mu_{ij}}^{m}\norm{x_{j}-\upsilon_{i}}^{2} \geq 1}\\
    {\scriptsize
    \upsilon_{i} \geq 0,~i=1,...,c}\\
    {\scriptsize
    \sum_{i=1}^{c}{\upsilon_{i}} \geq 1.}
\end{gathered}  
\end{equation}
\end{subequations}
Thus the optimized value of the equation \eqref{9} will be presented by,
\begin{equation}\label{10}
    {\scriptsize
    p^{*}=inf\Big\{ \sum_{j=1}^{N}\sum_{i=1}^{c}{\mu_{ij}}^{m}\norm{x_{j}-\upsilon_{i}}^{2}|J_{E} \geq 1, \upsilon \in \Upsilon \Big\}.}
\end{equation}
However, the equations \eqref{9} and \eqref{10} have multiple sets of constraints including inequalities. As such directly applying ADMM optimization technique to solve the cost function will rise the mathematical complexity \cite{Ye2014ScalingUD}. So, to solve optimization problem \eqref{10}, it has been reformulated as a three-block separable form with linear equality constraints \cite{he2021extensions}:
\begin{equation}\label{11}
{\scriptsize
\begin{aligned}
    p^{*}=inf \Big\{ \sum_{j=1}^{N}\sum_{i=1}^{c}{\mu_{ij}}^{m}\norm{x_{j}-\upsilon_{i}}^{2} & |J_{E}-z=1,\\
    & \upsilon \in \Upsilon, z \in \mathfrak{{R_{+}}^{n}} \Big\},
\end{aligned}}
\end{equation}
where, $ z \in \mathfrak{{R_{+}}^{n}} $ denotes an auxiliary variables. Now, unlike the method of multipliers, here in the work the scaled augmented Lagrangian is formed as shown in equation \eqref{12}:
\begin{equation}\label{12}
{\scriptsize
\begin{aligned}
    & L_{\rho}(\mu, z, y, \upsilon, \lambda, \eta)= \sum_{j=1}^{N}\sum_{i=1}^{c}{\mu_{ij}}^{m}\norm{x_{j}-\upsilon_{i}}^{2}\\
    & + \rho\sum_{i,j}\lambda_{i,j}\Big( \sum_{j=1}^{N}\sum_{i=1}^{c}{\mu_{ij}}^{m}\norm{x_{j}-\upsilon_{i}}^{2}-z-1\Big)\\
    & + (\rho/2)\sum_{i,j}\Big( \sum_{j=1}^{N}\sum_{i=1}^{c}{\mu_{ij}}^{m}\norm{x_{j}-\upsilon_{i}}^{2}-z-1\Big)^{2}\\
    & + \tau\rho\eta\Big( \sum_{i=1}^{c}{\upsilon_{i}}-y-1 \Big) + \tau(\rho/2)\Big( \sum_{i=1}^{c}{\upsilon_{i}}-y-1 \Big)^{2}.
\end{aligned}}
\end{equation}
Here in equation \eqref{12}, $ \tau $ denotes the penalty parameter and the term $ \rho $ used for balancing the cost function and augmented Lagrangians. 

The 3-ADMM-H optimizer split the problem \eqref{12} into QUBO and convex optimization problems. Later the first block of 3-ADMM-H is used for processing QUBO part of the equation. The second and third block of 3-ADMM-H is used to update the convex and combination of convex and quadratic problems respectively. Here in this work the QUBO subproblem have been solved on the quantum device via Quantum Approximate Optimization Algorithm (QAOA). The QAOA is a variational quantum algorithms (VQA) use for solving the combinatorial optimization problems \cite{farhi2014quantum}. QAOA with $ k $ bit input basically depends on the integer parameter $ P $. The quantum algorithm find string $ \mathfrak{z} $ for which the cost function $ C(\mathfrak{z}) $ resembling the sum of $ M $ local terms, is approximately equivalent to $ C $'s global optimum. In general, for each call quantum computer utilize a set of $ 2P $ angles $ (\gamma, \beta) $ and produce the state as,
\begin{equation}\label{13}
{\scriptsize
    | \gamma, \beta\rangle = U(B,\beta_{P})U(C,\gamma_{P})...U(B,\beta_{1})U(C,\gamma_{1}) |s \rangle;} 
\end{equation}
where $ U(C,\gamma) $ is a unitary operator depends on angle $ \gamma $ and is represented by:
\begin{equation}\label{14}
{\scriptsize
    U(C,\gamma)=e^{-\mathfrak{i}\gamma C}= \prod_{\alpha=1}^{M}e^{-\mathfrak{i}\gamma C_{\alpha}},}
\end{equation}
and the term $ U(B,\beta) $ is given by,
\begin{equation}\label{15}
{\scriptsize
    U(B,\beta)=e^{-\mathfrak{i}\beta B}= \prod_{t=1}^{k}e^{-\mathfrak{i}\beta \sigma^{X}_{t}}.}
\end{equation}summarized
Here in equations \eqref{14} and \eqref{15} the term $ \gamma $ and $ \beta $ lies between $ 0 $ and $ 2\pi $, and $ 0 $ and $ \pi $ respectively. The term $ B $ represent the sum of all single bit $ \sigma^{X} $ operators denoted by,
\begin{equation}\label{16}
{\scriptsize
    B=\sum_{t}^{k}\sigma^{X}_{t}.}
\end{equation}
The notion $ |s\rangle $ denotes the initial state which will be uniform superposition over computational basis states \cite{farhi2014quantum}:
\begin{equation}\label{17}
{\scriptsize
    |s\rangle=\frac{1}{\sqrt{2^{k}}}\sum_{\mathfrak{z}}|\mathfrak{z} \rangle.}
\end{equation}
The QAOA has shown its capability in terms of performance with respect to the classical Goemans-Williamson limit \cite{crooks2018performance}. Alongside, the combination of multi-start strategies and hyperparametrization in QAOA has shown promising results in
escaping local optima \cite{8916288}. 

\begin{figure*}[ht]
 \tikzstyle{process} = [rectangle, minimum width=2cm, minimum height=1cm, text centered, draw=black]
	 	 	 \tikzstyle{process1} = [rectangle, minimum width=2cm, minimum height=2cm, text centered, draw=black]
	 	 	 \tikzstyle{process2} = [rectangle, minimum width=4cm, minimum height=1cm, text centered, draw=black]
	 	 	 \tikzstyle{process3} = [rectangle, minimum width=3cm, minimum height=0.5cm, text centered, draw=black]
	 	 	 \tikzstyle{naveqs} = [sensor, text width=6em, fill=red!20,minimum height=10em, rounded corners]
	 	 	 \tikzstyle{arrow} = [thick,->,>=stealth]
             \resizebox {\textwidth} {!} {
	 	 	 \begin{tikzpicture}[node distance=2cm]
	 	 	 	 \node (in1) [process,text width=1.6cm, xshift=0cm] {Solar Disk Image};
	 	 	 	 \node (in2) [process1, text width=2cm, right of=in1, yshift=0cm, xshift=0.5cm] {Generate the histogram of the image};
	 	 	 	 \node (pro1) [process1, text width=2.5cm, right of=in2, yshift=0cm, xshift=0.9cm] {Randomly initialize the membership values, threshold $ \epsilon $};
	 	 	 	 \node (pro2) [process1, text width=2cm, right of=pro1, yshift=0cm, xshift=0.8cm] {Minimize $ J_{E} $ using M-ADMM-H};
                 \node (pro21) [process1, text width=2.2cm, right of=pro2, yshift=0cm, xshift=0.7cm] {Calculate cluster center $\upsilon_{i}$ such that $\upsilon_{1},..., \upsilon_{i}=arg \min\limits_{1\leq i\leq c}J_{E}$};       
	 	 	 	 \node (pro51) [process1, text width=2.2cm, right of=pro21, yshift=0cm, xshift=0.9cm] {Update membership value using Equation (7)};
	 	 	 	 \node (pro6) [process1, text width=2cm, right of=pro51, xshift=0.8cm] {Compare $ J_{E} $ with respect to $ \epsilon $};
	 	 	 	 \node (pro7) [process1, text width=2cm, right of=pro6, yshift=0cm, xshift=2.1cm] {Cluster image into foreground and background};
	 	 	 	 \node (pro8) [process1, text width=2cm, right of=pro7, yshift=0cm, xshift=1.5cm] {Initialize circular contour};
	 	 	 	 \node (pro10) [process3, text width=3cm, below of=pro8, yshift=-1.5cm, xshift=-1.5cm] {Coronal holes regions};
	 	 	 	 \node (pro12) [process3, text width=3cm, above of=pro8, yshift=3cm, xshift=-1.6cm] {Non-Coronal holes regions};
	 	 	 	 
	 	 	 	 \draw [arrow] (in1) -- (in2);
	 	 	 	 \draw [arrow] (in2) -- (pro1);
	 	 		 \draw [arrow] (pro1.east)-++(0.3,0)|- (pro2);
	 	 		 \draw [arrow] (pro2) -- (pro21);
                          \draw [arrow] (pro21) -- (pro51);
	 	 		 \draw [arrow] (pro51) -- (pro6);
	 	 		 \draw [-stealth] (pro6.east)-++(0,0)|- (pro7)node[yshift=0.3cm, xshift=-0.7cm,text width=4cm]{If $ J_{E}<\epsilon $};
	 	 		 \draw [arrow] (pro6.north)-- ++(0,1.4)node[xshift=-3.5cm,pos=1.2,text width=3cm]{If $ J_{E}>\epsilon $}-|++(-6.1,0)-| (pro2.north);
	 	 		 
	 	 		 \draw [-stealth] (pro8.south)--++(0.0,-1.5)-| (pro10.north)node[yshift=1.0cm, xshift=-1.8cm,text width=4cm]{foreground};
	 	 		 \draw [-stealth] (pro7.south)--++(0.0,-1.38)-| (pro10.north)node[yshift=1.2cm, xshift=2.6cm,text width=2.5cm]{\begin{center}pixels inside contour \end{center}};
	 	 		 
	 	 		 \draw [-stealth] ($ (pro8.north)+ (0.1,0) $)--++(0,1.5)--++(-4.31,0)--++(0,0.9)|-($ (pro12.west)+ (0,0) $)node[yshift=-3.5cm, xshift=-0.7cm,text width=4cm]{foreground};
	 	 		 \draw [-stealth] ($ (pro7.north)+ (-0.72,0) $)--++(0,1.3)|- ($ (pro12.west) + (0,0) $)node[yshift=-1.8cm, xshift=0.4cm,text width=2.5cm]{\begin{center}pixels outside contour \end{center}};
	 	 		 
	 	 		 \draw [-stealth] ($ (pro8.north)+ (0.8,0) $)--++(0,1.3)|- ($ (pro12.east)+ (0,0) $)node[yshift=-3.7cm, xshift=-1.1cm,text width=4cm]{background};
	 	 		 \draw [-stealth] ($ (pro7.north)+ (0.2,0) $)--++(0,0.8)--++(4.1,0)--++(0,0.9)|-($ (pro12.east) + (0,0) $)node[yshift=-3.3cm, xshift=2.0cm,text width=2.5cm]{\begin{center}pixels inside contour \end{center}};

	 	 	 \end{tikzpicture}
             }
	\caption{Block-diagram of the proposed FFCM-SCM based coronal holes detection in solar disk image.}\label{Block-diagram_2}
\end{figure*}
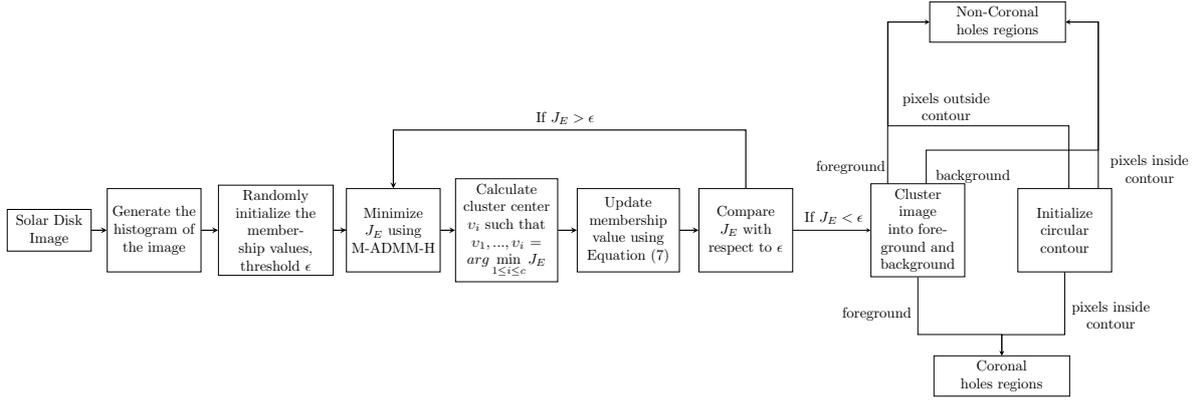

After the clustering image based on the optimal cluster center obtained using quantum-FFCM, into two regions, namely, foreground regions consisting of CHs regions and solar image background, and background regions consisting of non-CHs regions. A static contour is initialized using circular Hough transform based on the original solar disk image. The contour has been initialized with a motive to extract out the CHs regions from the segmented solar image and to separate out the redundant solar limb. Based on three conditions as specified underneath the proposed model can effectively distinguish the CHs. 
\begin{itemize}
	\item[$ \bullet $] In the event that the foreground region is present exterior of the circular contour, set as background.
	\item[$ \bullet $] Background region is present interior of the contour, set as background.
	\item[$ \bullet $] The foreground region is exists interior of the circular contour, set as foreground CHs regions.
\end{itemize} 
The overall block diagram of the proposed technique of QFFCM based CHs detection has been shown in Fig. \ref{Block-diagram_2}. 

\section{Results and Discussions}\label{Section-4}
\begin{figure}
    \centering
    \includegraphics[width=0.95\linewidth]{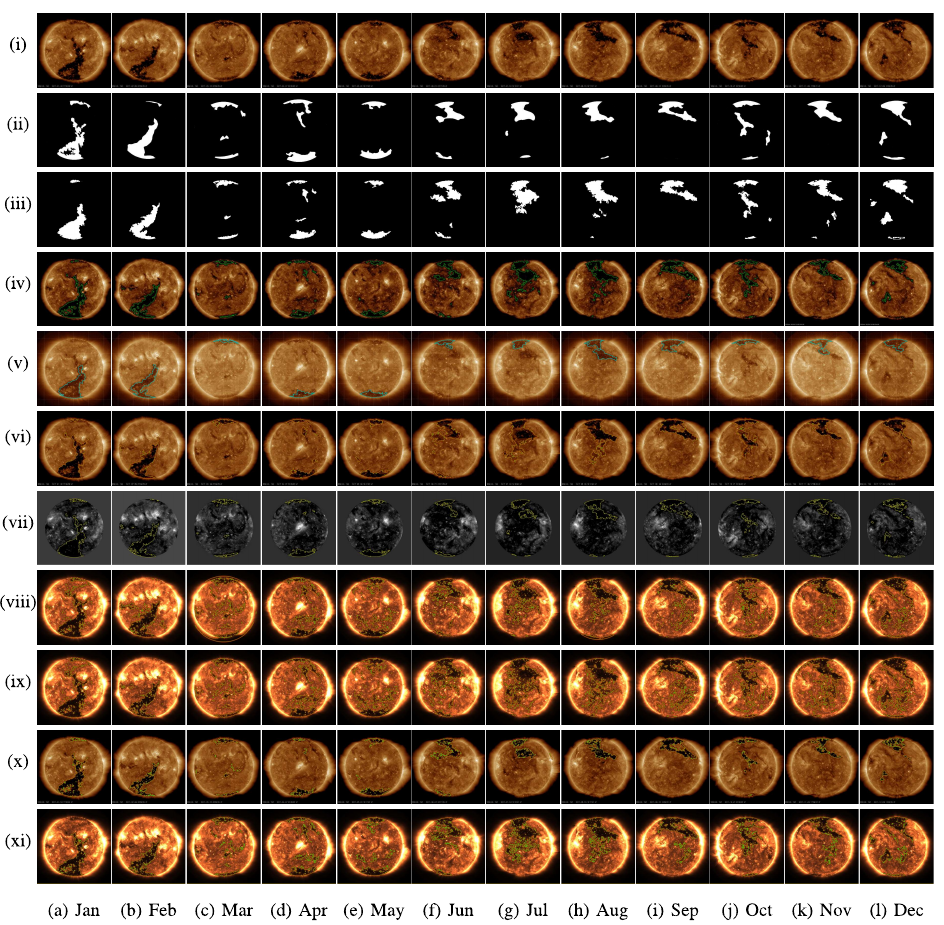}
    \caption{A pictorial summarization showing (i) Original Image followed by ground-truth images in (ii) and (iii), and result obtained using model like, CNN Seg, SPoCA, CHIMERA, ACWE, FFCM-SCM, POR-ACM, FEDCM and the proposed method in (iv)-(xi) respectively.}
    \label{Fig 1.}
\end{figure}
For experimental purpose in this work one EUV band pass image of 193\AA{}, per day for the year 2017 are used to detect CHs. These solar images had been captured using Atmospheric Imaging Assembly on-board Solar Dynamics Observatory (SDO/AIA) centered on the specific line of 193 \AA{} \cite{2012SoPh..275....3P}. These SDO/AIA images have been collected from both Joint Science Operations (JSOC) and Solar Dynamics Observatory (SDO) portal. However, in this work all the experiment have been carried out based on the images collected from JSOC and the images collected from SDO portal is only for the visualization purpose.

The experiment has been carried out by setting fuzziness parameter, $m$ equal to 2 ($>$1), as for $1\leftarrow m$, the model will become equivalent to classical k-means clustering technique, whereas for higher value of $m$ all pixels will have identical membership to each cluster. Here, the performance of the proposed method, QFFCM have been analyze in terms of quality, quantity and execution time requirement with the results obtained from both existing supervised and unsupervised methods such as ACWE, SPoCA, CNN, POR-ACM, FFCM-SCM, FEDCM and coronal hole identification via multi-thermal emission recognition algorithm (CHIMERA) methods. The outcomes have been tally with two ground-truth images generated based on the synoptic map image provided by NOAA\textbackslash SWPC (SM-GT) and region-growing based CHs segmentation map \footnote{\url{https://github.com/observethesun/coronal_holes/tree/master/data}} (RG-GT). The performance of the QFFCM have been analyze in terms of quality, quantity and execution time requirement. 

\subsection{Visual Analysis}
Based on visual analysis, it can be concluded that the proposed QFFCM approach for detecting CHs exhibits comparable performance to other existing methods of CHs detection. When visually comparing the outcomes obtained using the SPoCA, CHIMERA, and ACWE methods, shown in the first column of Figs. 2v - 2vii respectively, it can be inferred that the proposed QFFCM-based approach for CHs detection has identified the CHs region more accurately than these methods, similar to SM-GT. It is evident that the existing SPoCA, CHIMERA, and ACWE methods have incorrectly classified non-CHs regions as CHs regions for the solar image captured in January 2017. However, when comparing the outcomes of the proposed method with RG-GT, it can be noted that the proposed method has misclassified some regions as non-CHs, which are not present in RG-GT. It is noteworthy that RG-GT has been generated using a region-growing based method, whose accuracy depends entirely on the proper selection of initial seed points in the image. Additionally, when comparing the outcomes of the proposed method with those of CNN Seg shown in the first column of Fig. 2iv, it can be observed that the QFFCM-based approach outperforms CNN Seg. In comparison with POR-ACM, FFCM-SCM, and FEDCM, the QFFCM-based approach shows near-similar performance.

For the case of Fig. 2b, it has been observed that the QFFCM-based approach of CHs detection is capable of identifying most of the curvatures associated with the CHs boundary, as can be seen in the second column of Fig. 2iii. In the same case, other compared techniques are unable to identify these curvatures. It is also worth mentioning that the proposed technique is not as suitable when compared with SM-GT, as SM-GT is a hand-drawn generated ground-truth where locating every change in curvature associated with the CHs region is quite a difficult task.

For the solar image captured in March 2017, shown in Fig. 2c, it can be visually observed that the methods CNN Seg, SPoCA, CHIMERA, and ACWE are unable to locate the CHs regions properly. However, for the case of POR-ACM, it can be seen that the method has misclassified redundant regions as CHs regions. When comparing the outcome of FFCM-SCM with that of QFFCM, it can be concluded that the proposed method is capable of providing similar results to FFCM-SCM, as both methods are based on fuzzy c-mean techniques.

Throughout the visual analysis, it has been noticed that in most cases, the proposed technique has detected the actual CHs regions in the solar images, but it has also misclassified some redundant regions as CHs regions. The main reason behind this performance of the proposed method is that the QFFCM method used in the initial stage of CHs detection essentially aims to segment the solar image into two parts based on the intensity level of image pixels: darker regions and non-darker regions. Throughout the experiment, it has been observed that the method is efficient in extracting all darker regions from the images. However, the next step of the proposed work depends on area-based image morphological operations, where redundant regions in the segmented image are removed based on the area of the segmented part. This means that if the area of the segmented regions is greater than a certain threshold value, it should be considered as CHs regions; otherwise, they are considered non-CHs regions. Choosing this area threshold value is a vital task in this work. Selecting a higher value for the area threshold may lead to the elimination of the region of interest, whereas the opposite may lead to the selection of redundant regions. Therefore, one must be optimal in choosing this value. In the future, it will be a focus to choose the optimal value of the area threshold for more proper and accurate detection of CHs regions.   
\begin{figure}[!h]
	\centering
	{\includegraphics[scale=1.2]{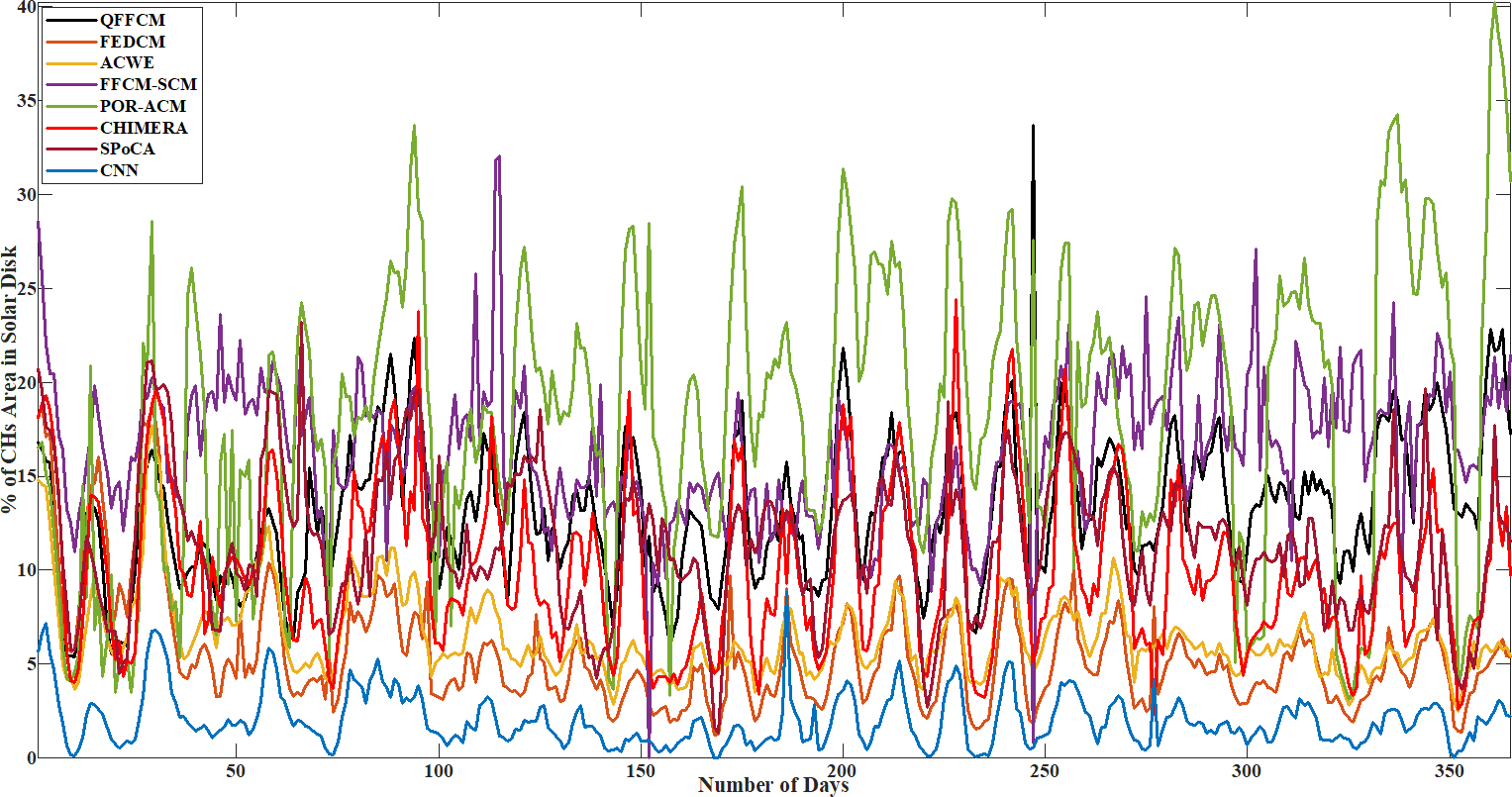}}
	\caption{CHs area detected for the year \textit{2017}, using existing methods and proposed method QFFCM.}\label{Fig 2.}	
\end{figure} 

\begin{figure}[!h]
	\centering
	{\includegraphics[scale=1.2]{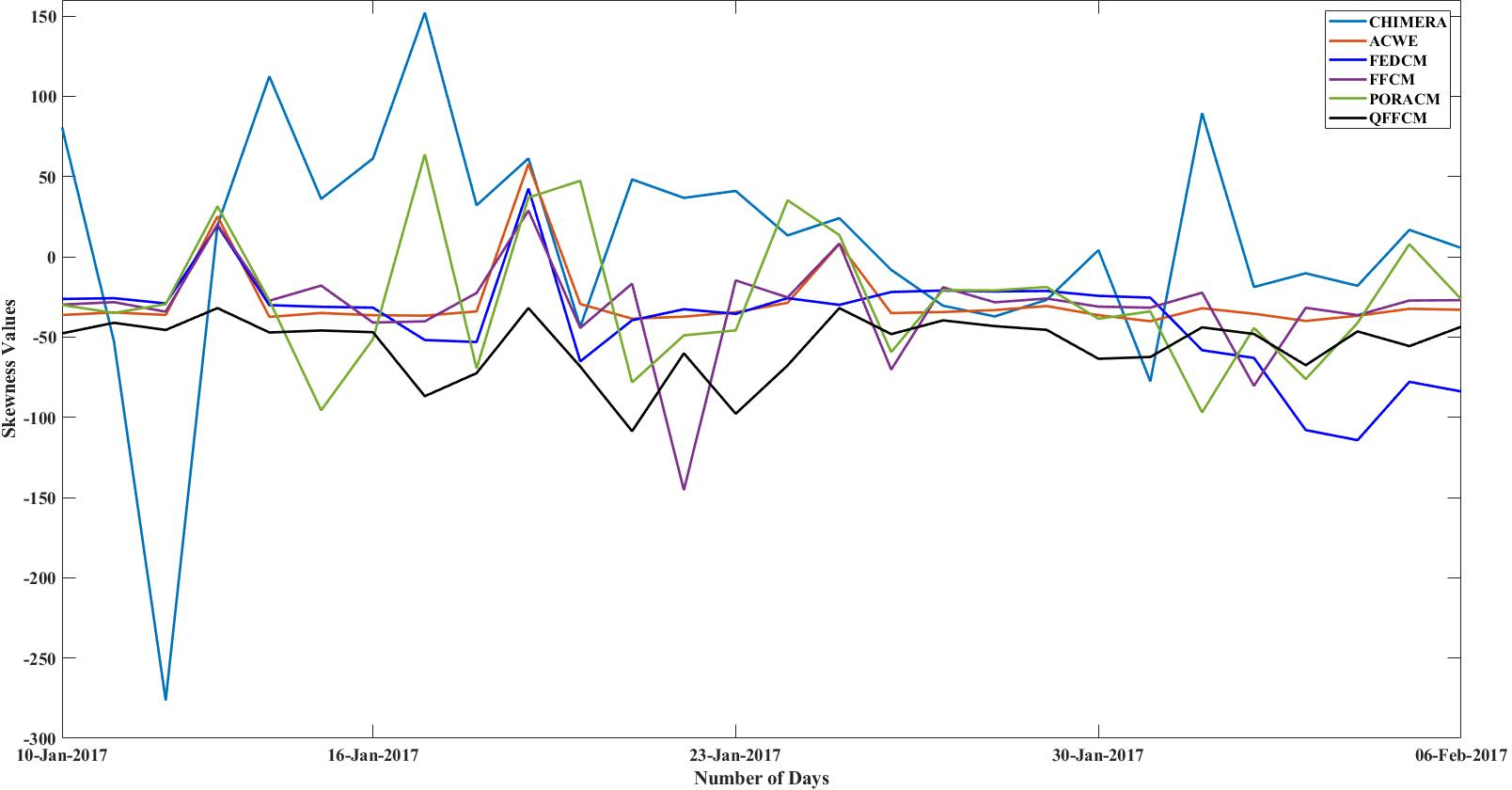}}
	\caption{Skewness value detected for a carrington rotation in the year \textit{2017}, using existing methods and proposed method QFFCM.}\label{Fig 3.}	
\end{figure}

\subsection{Quantitative Analysis}
\begin{sidewaysfigure}
    \centering
    \includegraphics[width=0.70\linewidth, angle=-90]{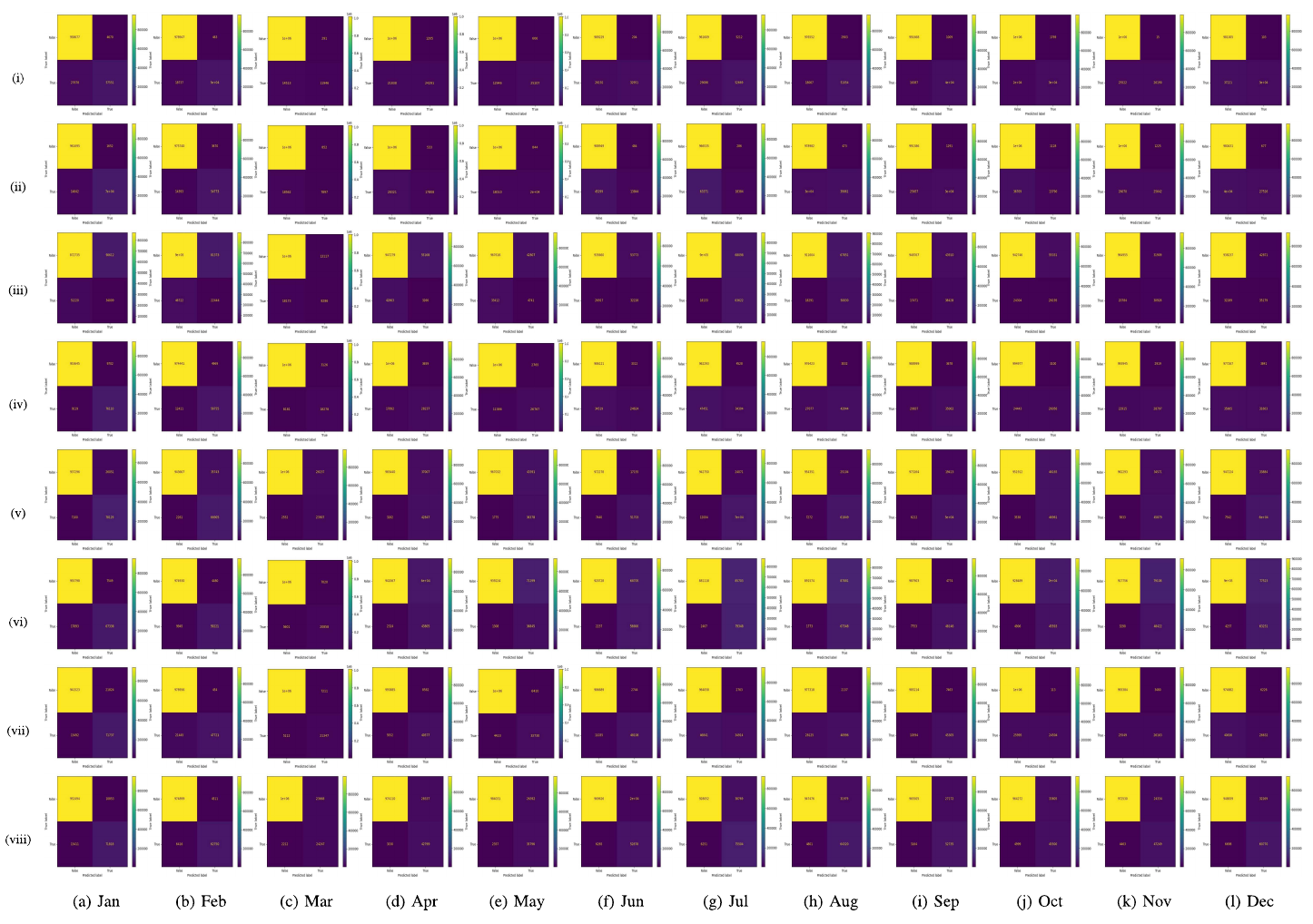}
    \caption{A pictorial summary depicting the confusion matrix generated for results obtained using models such as CNN Seg, SPoCA, CHIMERA, ACWE, FFCM-SCM, POR-ACM, FEDCM, and the proposed method in (i)-(viii) respectively, when compared with the SM-GT.}
    \label{Fig 4.}
\end{sidewaysfigure}

\begin{sidewaysfigure}
    \centering
    \includegraphics[width=0.70\linewidth, angle=-90]{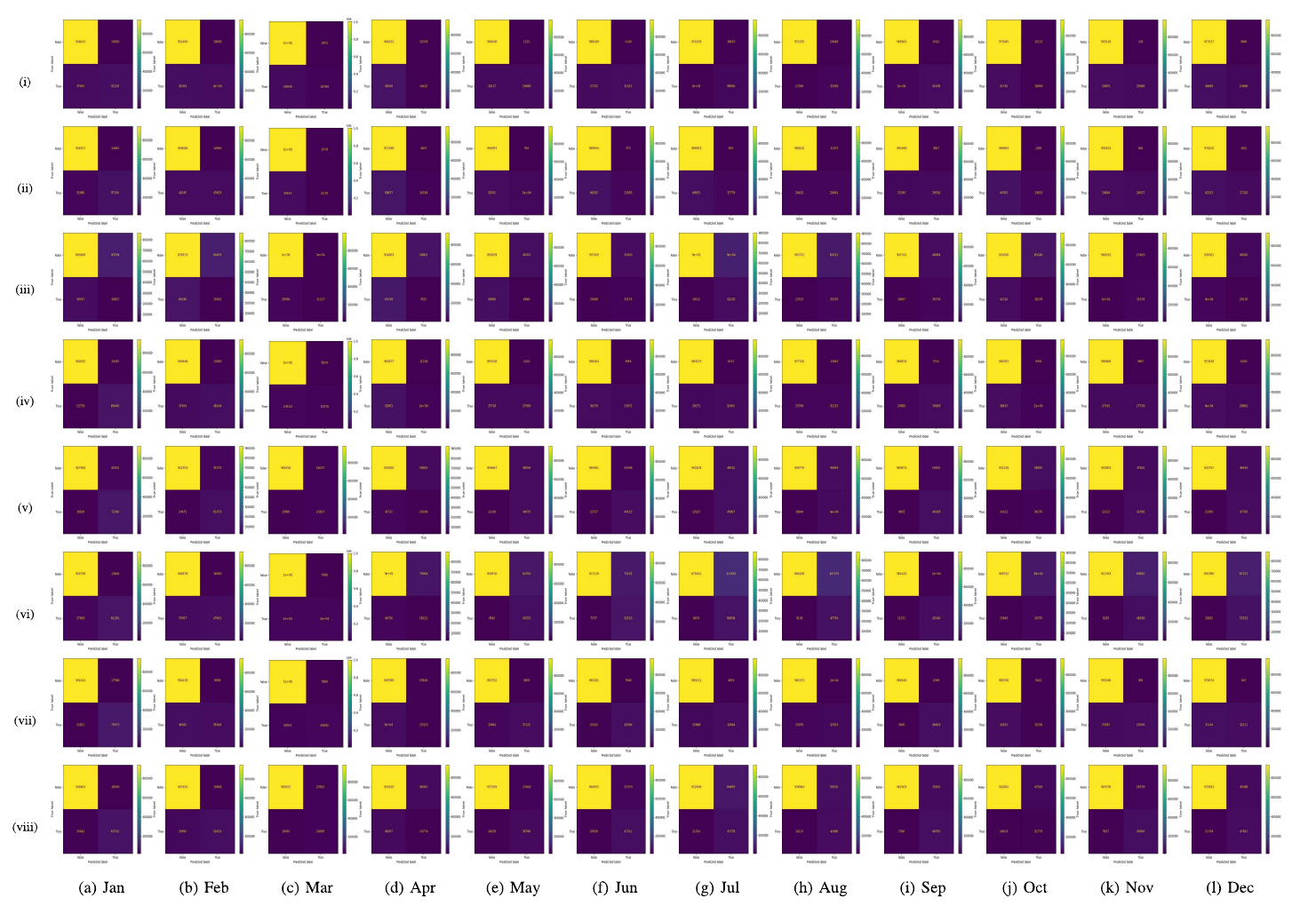}
    \caption{A pictorial summary depicting the confusion matrix generated for results obtained using models such as CNN Seg, SPoCA, CHIMERA, ACWE, FFCM-SCM, POR-ACM, FEDCM, and the proposed method in (i)-(viii) respectively, when compared with the RG-GT.}
    \label{Fig 5.}
\end{sidewaysfigure}

For quantitative analysis F1 score (shown in Table \ref{tab1}) given by,
\begin{equation}\label{18}
{ \begin{split}
	F1~score = \dfrac{2|A\bigcap B|}{|A|+|B|},
	\end{split} } 
\end{equation}   
accuracy rate (shown in Table \ref{tab2}) denoted by,  
\begin{equation}\label{19}
{ \begin{split}
	Accuracy~rate & = \dfrac{|A|}{|B|};\\
	\end{split} } 
\end{equation}  
total area covered (plotted in Fig. \ref{Fig 2.}) and magnetic skewness value (plotted in Fig. \ref{Fig 3.}) have been considered. Where, in equations \eqref{18} and \eqref{19}, the term $ A $ denotes output image or region inside the final curve and $ B $ represents the ground-truth of solar image. This F1 score and accuracy rate have been calculated based on the values of confusion matrix shown in Figs. \ref{Fig 4.} and \ref{Fig 5.}. Now, comparing the outcome of the QFFCM with other existing methodologies and two ground-truth images, it can be stated that the performance of the method is comparable with that of CHIMERA and SPoCA. For the solar image captured on \textit{Feb} shown in Fig. 2b, while compared with the RG-GT, the proposed QFFCM detect CHs with heighest F1 score of 91.99$\%$, as can be seen in Table \ref{tab1}. Since, in this case the QFFCM hasn't detected the northern polar coronal hole region similar to RG-GT. However, comparing with SM-GT and RG-GT it can be stated that, in rest of the the cases the QFFCM have detected some non-CHs regions. This may be due to the improper selection of parameter in redundant pixel removal algorithm or improper assignment of the pixel value while transforming the solar FITS file image of 32-bit to 8-bit values. Also, it should be noted that the degradation in the F1 score and Accuracy Measure values for the outcomes obtained using the proposed QFFCM-based approach for CHs detection might be attributed to the calculation of these values based on an improper ground-truth dataset. Upon visual comparison between the outcomes obtained using the proposed technique and the utilized ground-truth, it can be observed that the proposed technique has detected very subtle changes in the curvature of CH boundaries that were not identified in the provided ground-truth. Additionally, the QFFCM-based approach for CH detection has identified many tiny non-CHs regions that were not indicated in the ground-truth dataset. Now, while comparing the performance of QFFCM in terms of total area covered by the detected CHs and magnetic skewness of the extracted CHs, it has been found that the line graph for the proposed method follow the similar trends as that of the existing one.  

\subsection{Execution Time Analysis}
All the algorithms have been executed on a system having configured with 4GB RAM and Intel(R) Core(TM)i5-8250U CPU, having the base speed of 1.8GHz, along-with Windows 10, 64-bit operating system. Based on this the execution time requirement of all the algorithms have been summarized in the Table \ref{tab6-2}, and it has been found that the QFFCM generates the CHs segmentation results within 12 sec, which is near to real-time. This is due to the incorporation of the quantum computing based optimization technique which provide very faster output compared to classical computing methods. 
\begin{table*}[!ht]
	{\renewcommand\arraystretch{1.50}
		\caption{Coronal Holes' F1 Score based on SM-GT and RG-GT.}\label{tab1}
		\resizebox{\textwidth}{!}
		{\centering\small
			\begin{tabular}{|>{\centering\arraybackslash}m{1.8cm}|>{\centering\arraybackslash}m{1.5cm}|c|c|c|c|c|c|c|c|c|c|c|c|}
				\hline
				\multirow{2}{*}{Methods}&\multirow{2}{*}{\begin{varwidth}[t]{\linewidth} \begin{center}	Ground-truth\\ Used \end{center} \end{varwidth}}&\multicolumn{12}{c|}{Solar Image Captured on}\\
				\cline{3-14}
				&  & Jan & Feb & Mar & Apr & May & Jun & Jul & Aug & Sep & Oct & Nov & Dec\\
				\hline
				\multirow{2}{*}{ACWE Seg}& \multirow{1}{*}{SM-GT} & $ 0.7020  $ & $ 0.6550 $ & $ 0.5047 $ & $ 0.3895 $ & $ 0.6770 $ & $ 0.5658 $ & $ 0.6665 $ & $ 0.6100 $ & $ 0.7066 $ & $ 0.4603 $ & $ 0.6493 $ & $ 0.5636 $ \\
				\cline{2-14}
				& \multirow{1}{*}{RG-GT} & $0.8899$ & $0.8672$ & $0.7637$ & $0.7237$ & $0.7909$ & $0.5655$ & $0.5689$ & $0.7363$ & $0.7409$ & $0.6542$ & $0.6903$ & $0.6141$\\
				\hline
				\multirow{2}{*}{CNN Seg}& \multirow{1}{*}{SM-GT} & $0.6768$ & $0.5901$ & $0.4150$ & $0.2929$ & $0.6279$ & $0.6786$ & $0.6700$ & $0.6009$ & $0.7245$ & $0.3693$ & $0.6425$ & $0.4947$ \\
				\cline{2-14}
				& \multirow{1}{*}{RG-GT} & $0.7806$ & $0.8400$ & $0.6174$ & $0.6782$ & $0.7881$ & $0.7140$ & $0.7543$ & $\textbf{\textit{0.8296}}$ & $0.8195$ & $\textbf{\textit{0.7296}}$ & $0.6722$ & $0.6184$ \\
				\hline
				\multirow{2}{*}{SPoCA}& \multirow{1}{*}{SM-GT} & $ 0.7070 $ & $ 0.6043 $ & $ 0.2550 $ & $ 0.3212 $ & $ 0.5648 $ & $ 0.3960 $ & $0.4675$ & $0.5902$ & $0.6752$ & $0.3869$ & $0.6473$ & $0.5725$ \\
				\cline{2-14}
				& \multirow{1}{*}{RG-GT} & $\textbf{\textit{0.8949}}$ & $0.8583$ & $0.4484$ & $0.5529$ & $0.6811$ & $0.3771$ & $0.3660$ & $0.7235$ & $0.6916$ & $0.4215$ & $0.6527$ & $0.5750$ \\
				\hline
				\multirow{2}{*}{CHIMERA}& \multirow{1}{*}{SM-GT} & $0.8010$ & $\textbf{\textit{0.7377}}$ & $0.5195$ & $\textbf{\textit{0.4509}}$ & $0.7820$ & $0.7694$ & $0.6483$ & $0.6360$ & $0.8625$ & $\textbf{\textit{0.5386}}$ & $\textbf{\textit{0.8661}}$ & $\textbf{\textit{0.6932}}$ \\
				\cline{2-14}
				& \multirow{1}{*}{RG-GT} & $0.8473$ & $0.8332$ & $0.6788$ & $0.7160$ & $0.8188$ & $0.7706$ & $0.7004$ & $0.7744$ & $0.8737$ & $0.6802$ & $\textbf{\textit{0.8064}}$ & $\textbf{\textit{0.7948}}$ \\
				\hline	
				\multirow{2}{*}{\begin{varwidth}[t]{\linewidth} \begin{center}FFCM- SCM \end{center} \end{varwidth}} & \multirow{1}{*}{SM-GT} & $0.7473$ & $0.6489$ & $0.5308$ & $0.3828$ & $0.6135$ & $0.7004$ & $0.6004$ & $0.5568$ & $0.7199$ & $0.4510$ & $0.6293$ & $0.5804$ \\
				\cline{2-14}
				& \multirow{1}{*}{RG-GT} & $0.8249$ & $0.7788$ & $0.6248$ & $0.6802$ & $0.6169$ & $0.8078$ & $\textbf{\textit{0.7945}}$ & $0.7925$ & $0.7950$ & $0.6539$ & $0.6942$ & $0.7431$ \\
				\hline
				\multirow{2}{*}{\begin{varwidth}[t]{\linewidth} \begin{center}POR-ACM\end{center} \end{varwidth}} & \multirow{1}{*}{SM-GT} & $0.7471$ & $0.6384$ & $0.5890$ & $0.3186$ & $0.5377$ & $0.5712$ & $0.4526$ & $0.4502$ & $0.7966$ & $0.4001$ & $0.5112$ & $0.5094$ \\
				\cline{2-14}
				& \multirow{1}{*}{RG-GT} & $0.8411$ & $0.8914$ & $0.7677$ & $0.5828$ & $0.5040$ & $0.6160$ & $0.6430$ & $0.6003$ & $\textbf{\textit{0.8848}}$ & $0.5503$ & $0.5403$ & $0.6073$ \\
				\hline				
				\multirow{2}{*}{FEDCM}& \multirow{1}{*}{SM-GT} & $ \textbf{\textit{0.8247}} $ & $ 0.5905 $ & $ \textbf{\textit{0.6029}} $ & $ 0.3810 $ & $ \textbf{\textit{0.8137}} $ & $ \textbf{\textit{0.7850}} $ & $ \textbf{\textit{0.6873}} $ & $ \textbf{\textit{0.6569}} $ & $ \textbf{\textit{0.8671}} $ & $ 0.3719 $ & $ 0.7063 $ & $ 0.6470 $ \\
				\cline{2-14}
				& \multirow{1}{*}{RG-GT} & $0.8024$ & $0.8133$ & $\textbf{\textit{0.7760}}$ & $\textbf{\textit{0.8518}}$ & $\textbf{\textit{0.8616}}$ & $\textbf{\textit{0.8747}}$ & $0.5846$ & $0.7304$ & $0.8338$ & $0.6524$ & $0.6431$ & $0.5309$ \\
				\hline
				\multirow{2}{*}{QFFCM}& \multirow{1}{*}{SM-GT} & $0.7418$ & $0.6759$ & $0.5512$ & $0.3444$ & $0.6755$ & $0.7228$ & $0.5481$ & $0.5576$ & $0.6986$ & $0.4620$ & $0.7122$ & $0.5869$ \\
				\cline{2-14}
				& \multirow{1}{*}{RG-GT} & $0.8555$ & $\textbf{\textit{0.9199}}$ & $0.6517$ & $0.7292$ & $0.7128$ & $0.8022$ & $0.7704$ & $0.7777$ & $0.7767$ & $0.7008$ & $0.7663$ & $0.7572$ \\
				\hline								
		\end{tabular}}
	}
\end{table*}

\begin{table*}[!ht]
	{\renewcommand\arraystretch{1.50}
		\caption{Coronal Holes' Accuracy Measure based on SM-GT and RG-GT.}\label{tab2}
		\resizebox{\textwidth}{!}
		{\centering\small
			\begin{tabular}{|>{\centering\arraybackslash}m{1.8cm}|>{\centering\arraybackslash}m{1.5cm}|c|c|c|c|c|c|c|c|c|c|c|c|}
				\hline
				\multirow{2}{*}{Methods}&\multirow{2}{*}{\begin{varwidth}[t]{\linewidth} \begin{center}	Ground-truth\\ Used \end{center} \end{varwidth}}&\multicolumn{12}{c|}{Solar Image Captured on}\\
				\cline{3-14}
				&  & Jan & Feb & Mar & Apr & May & Jun & Jul & Aug & Sep & Oct & Nov & Dec\\
				\hline
				\multirow{2}{*}{ACWE Seg}& \multirow{1}{*}{SM-GT} & $0.9599$ & $0.9514$ & $0.9714$ & $0.9385$ & $0.9732$ & $0.9635$ & $0.9689$ & $0.9621$ & $0.9746$ & $0.9542$ & $0.9703$ & $0.9551$ \\
				\cline{2-14}
				& \multirow{1}{*}{RG-GT} & $0.9821$ & $0.9834$ & $0.9892$ & $0.9795$ & $0.9865$ & $0.9639$ & $0.9504$ & $0.9713$ & $0.9766$ & $0.9737$ & $0.9754$ & $0.9621$\\
				\hline
				\multirow{2}{*}{CNN Seg}& \multirow{1}{*}{SM-GT} & $0.9521$ & $0.9470$ & $0.9709$ & $0.9347$ & $0.9735$ & $0.9721$ & $0.9652$ & $0.9593$ & $0.9769$ & $0.9462$ & $0.9715$ & $0.9537$ \\
				\cline{2-14}
				& \multirow{1}{*}{RG-GT} & $0.9768$ & $0.9823$ & $0.9855$ & $0.9779$ & $0.9884$ & $0.9743$ & $\textbf{\textit{0.9679}}$ & $\textbf{\textit{0.9806}}$ & $0.9847$ & $\textbf{\textit{0.9778}}$ & $0.9748$ & $0.9651$ \\
				\hline
				\multirow{2}{*}{SPoCA}& \multirow{1}{*}{SM-GT} & $0.9554$ & $0.9457$ & $0.9655$ & $\textbf{\textit{0.9402}}$ & $0.9682$ & $0.9574$ & $0.9601$ & $0.9624$ & $0.9730$ & $\textbf{\textit{0.9561}}$ & $0.9717$ & $0.9588$ \\
				\cline{2-14}
				& \multirow{1}{*}{RG-GT} & $\textbf{\textit{0.9843}}$ & $0.9828$ & $0.9815$ & $0.9725$ & $0.9820$ & $0.9564$ & $0.9393$ & $0.9712$ & $0.9743$ & $0.9639$ & $0.9740$ & $0.9612$ \\
				\hline
				\multirow{2}{*}{CHIMERA}& \multirow{1}{*}{SM-GT} & $0.9641$ & $\textbf{\textit{0.9579}}$ & $0.9735$ & $0.9362$ & $0.9807$ & $0.9726$ & $0.9471$ & $0.9493$ & $0.9848$ & $0.9462$ & $\textbf{\textit{0.9870}}$ & $\textbf{\textit{0.9623}}$ \\
				\cline{2-14}
				& \multirow{1}{*}{RG-GT} & $0.9730$ & $0.9758$ & $0.9864$ & $0.9743$ & $0.9865$ & $0.9725$ & $0.9484$ & $0.9660$ & $0.9858$ & $0.9651$ & $\textbf{\textit{0.9817}}$ & $\textbf{\textit{0.9752}}$ \\
				\hline	
				\multirow{2}{*}{\begin{varwidth}[t]{\linewidth} \begin{center}FFCM- SCM \end{center} \end{varwidth}} & \multirow{1}{*}{SM-GT} & $0.9534$ & $0.9370$ & $0.9598$ & $0.9100$ & $0.9511$ & $0.9637$ & $0.9419$ & $0.9392$ & $0.9672$ & $0.9207$ & $0.9522$ & $0.9347$ \\
				\cline{2-14}
				& \multirow{1}{*}{RG-GT} & $0.9684$ & $0.9638$ & $0.9726$ & $0.9616$ & $0.9569$ & $0.9765$ & $0.9656$ & $0.9691$ & $0.9756$ & $0.9526$ & $0.9615$ & $0.9605$ \\
				\hline
				\multirow{2}{*}{\begin{varwidth}[t]{\linewidth} \begin{center}POR-ACM\end{center} \end{varwidth}} & \multirow{1}{*}{SM-GT} & $0.9604$ & $0.9486$ & $0.9735$ & $0.8849$ & $0.9291$ & $0.9248$ & $0.8832$ & $0.8888$ & $0.9793$ & $0.9007$ & $0.9150$ & $0.9017$ \\
				\cline{2-14}
				& \multirow{1}{*}{RG-GT} & $0.9757$ & $0.9862$ & $0.9880$ & $0.9402$ & $0.9309$ & $0.9323$ & $0.9160$ & $0.9145$ & $\textbf{\textit{0.9881}}$ & $0.9290$ & $0.9214$ & $0.9220$ \\
				\hline				
				\multirow{2}{*}{FEDCM}& \multirow{1}{*}{SM-GT} & $\textbf{\textit{0.9706}}$ & $0.9472$ & $\textbf{\textit{0.9744}}$ & $0.9280$ & $\textbf{\textit{0.9822}}$ & $\textbf{\textit{0.9778}}$ & $\textbf{\textit{0.9707}}$ & $\textbf{\textit{0.9674}}$ & $\textbf{\textit{0.9868}}$ & $0.9498$ & $0.9753$ & $0.9640$ \\
				\cline{2-14}
				& \multirow{1}{*}{RG-GT} & $0.9663$ & $0.9791$ & $\textbf{\textit{0.9883}}$ & $\textbf{\textit{0.9865}}$ & $\textbf{\textit{0.9897}}$ & $\textbf{\textit{0.9869}}$ & $0.9527$ & $0.9711$ & $0.9828$ & $0.9751$ & $0.9723$ & $0.9551$ \\
				\hline
				\multirow{2}{*}{QFFCM}& \multirow{1}{*}{SM-GT} & $0.9577$ & $0.9528$ & $0.9624$ & $0.9097$ & $0.9644$ & $0.9654$ & $0.9254$ & $0.9353$ & $0.9615$ & $0.9293$ & $0.9652$ & $0.9360$ \\
				\cline{2-14}
				& \multirow{1}{*}{RG-GT} & $0.9769$ & $\textbf{\textit{0.9896}}$ & $0.9752$ & $0.9697$ & $0.9725$ & $0.9751$ & $0.9571$ & $0.9649$ & $0.97105$ & $0.9629$ & $0.9725$ & $0.9628$ \\
				\hline								
		\end{tabular}}
	}
\end{table*}

\begin{table*}[h!]
	{\renewcommand\arraystretch{1.80}
	\vspace{0.5cm}
	\caption{Time Period and Iteration Required for Execution of Coronal Segmentation Technique.}\label{tab6-2}
		{\large
		\resizebox{\textwidth}{!}
		{\centering\small
			\begin{tabular}{|c|>{\centering\arraybackslash}m{2.cm}|c|c|c|c|c|c|c|c|c|c|c|c|c|c|c|}
				\hline
				\multirow{4}{*}{Images}&\multicolumn{16}{c|}{Methods}\\[0pt]
				\cline{2-17}
				&\multicolumn{2}{c|}{ACWE Seg}&\multicolumn{2}{c|}{FFCM-SCM}&\multicolumn{2}{c|}{POR-ACM}&\multicolumn{2}{c|}{CNN Seg}&\multicolumn{2}{c|}{SPoCA}&\multicolumn{2}{c|}{CHIMERA}&\multicolumn{2}{c|}{FEDCM}&\multicolumn{2}{c|}{QFFCM}\\[0pt]			
				
				&\multicolumn{2}{c|}{\begin{varwidth}[t]{\linewidth}\begin{center}(on Image size \\ $ 512\times512 $)\end{center} \end{varwidth}}&\multicolumn{2}{c|}{\begin{varwidth}[t]{\linewidth}\begin{center}(on Image size \\ $ 1024\times1024 $)\end{center} \end{varwidth}}&\multicolumn{2}{c|}{\begin{varwidth}[t]{\linewidth}\begin{center}(on Image size \\ $ 1024\times1024 $)\end{center} \end{varwidth}}&\multicolumn{2}{c|}{\begin{varwidth}[t]{\linewidth}\begin{center}(on Image size \\ $ 256\times256 $)\end{center} \end{varwidth}}&\multicolumn{2}{c|}{\begin{varwidth}[t]{\linewidth}\begin{center}(on Image size \\ $ 4096\times4096 $)\end{center} \end{varwidth}}&\multicolumn{2}{c|}{\begin{varwidth}[t]{\linewidth}\begin{center}(on Image size \\ $ 1024\times1024 $)\end{center} \end{varwidth}}&\multicolumn{2}{c|}{\begin{varwidth}[t]{\linewidth}\begin{center}(on Image size \\ $ 1024\times1024 $)\end{center} \end{varwidth}}&\multicolumn{2}{c|}{\begin{varwidth}[t]{\linewidth}\begin{center}(on Image size \\ $ 1024\times1024 $)\end{center} \end{varwidth}}\\[5pt]
				
				\cline{2-17}
				& Time (sec) & Iteration & Time (sec) & Iteration & Time (sec) & Iteration & Time (sec) & Iteration & Time (sec) & Iteration & Time (sec) & Iteration & Time (sec) & Iteration & Time (sec) & Iteration \\[0pt]
				\hline
				\textbf{(a)} Jan & 141.7941 & 3 & 57.9049 & \textbf{\textit{-}} & 71.9788 & 5 & 239.7257 & 357 & 52.2209 & 100 & 117.4300 & 25 & 58.290141 & 10 & \textbf{\textit{12.0000}} & 19 \\
				\hline
				\textbf{(b)} Feb & 140.5847 & 3 & 47.6665 & \textbf{\textit{-}} & 80.1936 & 6 & 241.6426 & 357 & 53.5869 & 100 & 89.1225 & 25 & 60.653845 & 10 & \textbf{\textit{12.5647}} & 19 \\
				\hline
				\textbf{(c)} Mar & 141.6941 & 3 & 45.6274 & \textbf{\textit{-}} & 98.7736 & 6 & 238.7274 & 357 & 41.9048 & 100 & 108.0168 & 25 & 59.345424 & 10 & \textbf{\textit{12.1238}} & 19 \\
				\hline
				\textbf{(d)} Apr & 139.8022 & 3 & 56.8507 & \textbf{\textit{-}} & 82.5312 & 5 & 239.0552 & 357 & 53.7006 & 100 & 68.5132 & 25 & 62.319634 & 10 & \textbf{\textit{12.8547}} & 19 \\
				\hline
				\textbf{(e)} May & 140.1534 & 3 & 49.4531 & \textbf{\textit{-}} & 86.7731 & 5 & 238.5822 & 357 & 49.4854 & 100 & 85.3057 & 25 & 55.057306 & 10 & \textbf{\textit{11.9861}} & 19 \\
				\hline
				\textbf{(f)} Jun & 139.7707 & 3 & 45.7722 & \textbf{\textit{-}} & 80.7622 & 5 & 238.5443 & 357 & 41.4631 & 100 & 114.9441 & 25 & 56.739033 & 10 & \textbf{\textit{12.3697}} & 19 \\
				\hline
				\textbf{(g)} Jul & 141.8857 & 3 & 56.5732 & \textbf{\textit{-}} & 96.1077 & 7 & 241.4772 & 357 & 44.1775 & 100 & 107.5324 & 25 & 59.356644 & 10 & \textbf{\textit{11.9921}} & 19 \\
				\hline
				\textbf{(h)} Aug & 139.2533 & 3 & 57.6962 & \textbf{\textit{-}} & 79.8691 & 5 & 240.3188 & 357 & 48.2032 & 100 & 117.5695 & 25 & 55.420662 & 10 & \textbf{\textit{12.4407}} & 19 \\
				\hline
				\textbf{(i)} Sep & 141.7132 & 3 & 51.9919 & \textbf{\textit{-}} & 99.1384 & 7 & 240.1994 & 357 & 54.3626 & 100 & 99.3444 & 25 & 54.804513 & 10 & \textbf{\textit{12.2533}} & 19 \\
				\hline
				\textbf{(j)} Oct & 141.6061 & 3 & 49.5830 & \textbf{\textit{-}} & 80.6480 & 6 & 238.5798 & 357 & 54.4733 & 100 & 62.1427 & 25 & 54.321401 & 10 & \textbf{\textit{12.1274}} & 19 \\
				\hline
				\textbf{(k)} Nov & 139.0139 & 3 & 44.6668 & \textbf{\textit{-}} & 79.2677 & 5 & 241.4121 & 357 & 42.3642 & 100 & 110.9478 & 25 & 58.301716 & 10 & \textbf{\textit{12.1135}} & 19 \\
				\hline
				\textbf{(l)} Dec & 140.1993 & 3 & 46.8701 & \textbf{\textit{-}} & 85.8925 & 6 & 240.4882 & 357 & 54.5589 & 100 & 116.0396 & 25 & 54.065932 & 10 & \textbf{\textit{12.5132}} & 19 \\
				\hline
		\end{tabular}}
	}}
\end{table*}

\section{Conclusion}\label{Section-5}
Here in the work quantum computing based fast fuzzy c-means (QFFCM) algorithm has been implemented for the fast detection of the solar CHs region. In the work quantum optimization methods have been applied to to obtained the optimal fuzzy cluster center for CHs detection. From the experiment it has been found that the method is capable of producing comparable output within near real-time execution period. In future proper measure to be taken in setting the pixels range of solar image and parameter of redundant pixels removal algorithm for obtain more accurate output.     

\section*{Acknowledgment}
The authors thankfully acknowledge the use of data courtesy of NASA/SDO and the AIA and HMI science teams. The authors also thankfully acknowledge NOAA\textbackslash SWPC.

\end{document}